\documentclass[10pt,notitlepage,nofootinbib,twocolumn]{revtex4-1}
\usepackage{graphicx,color}
\usepackage{hyperref}
\usepackage{amssymb}

\begin{document}

\title{Rapidity dependence of transverse-momentum multiplicity correlations}
\author{Adam Bzdak}
\email[E-Mail:]{bzdak@fis.agh.edu.pl}

\affiliation{AGH University of Science and Technology,\\
Faculty of Physics and Applied Computer Science,\\
30-059 Krak\'ow, Poland}

\begin{abstract}
Following previous work \cite{Bzdak:2012tp}, we propose to analyze the 
rapidity dependence of transverse momentum and transverse-momentum multiplicity correlations.
We demonstrate that the orthogonal polynomial expansion of the latter has the potential to discriminate between models of particle production.
\end{abstract}

\maketitle

\section{Introduction}
\label{sec: introduction}

One of the central problems in high-energy hadronic collisions
is to understand the longitudinal structure of systems created in
proton-proton (p+p), proton-nucleus (p+A) and nucleus-nucleus (A+A)
collisions.

Not long ago, it was argued that an event-by-event long-range fluctuation of
the fireball rapidity distribution results in rather peculiar two- and
multi-particle rapidity correlations \cite{Bzdak:2012tp,Bzdak:2015dja}.
Recent measurement by the ATLAS Collaboration at the LHC \cite%
{Aaboud:2016jnr} revealed new and rather unexpected scaling results on
asymmetric rapidity fluctuations in p+p, p+A and A+A interactions.
Recently, this problem has drawn a noticeable theoretical \cite%
{Jia:2015jga,Bozek:2015tca,Monnai:2015sca,Broniowski:2015oif,Schenke:2016ksl,Bzd-Dus,Ke:2016jrd,He:2017laa}
and experimental \cite{Aaboud:2016jnr,star:a1,alice:a1} interest, see also %
\cite%
{Bozek:2010vz,Bialas:2011xk,Bialas:2011bz,Csernai:2012mh,Pang:2012he,Vovchenko:2013viu,Olszewski:2015xba,Casalderrey-Solana:2013sxa,Vechernin:2015upa,Pang:2015zrq}
for recent related studies.

To summarize the main idea, the single-particle rapidity distribution in
each event, $N(y)$, can be written as
\begin{equation}
\frac{N(y)}{\left\langle N(y)\right\rangle }=1+a_{0}+a_{1}y+...,
\label{eq:N(y)}
\end{equation}%
where $a_{0}$ describes the rapidity independent fluctuation of the
fireball. $a_{1}$ represents the fluctuating long-range
forward-backward rapidity asymmetry.\footnote{%
We are not interested in statistical fluctuations, which can also generate
nonzero values of $a_i$. These are removed by measuring correlation
functions.} This coefficient can be driven for example by the difference in the
number of left- and right-going sources of particles, e.g., wounded nucleons 
\cite{Bialas:1976ed,Bialas:2004su}.  
$\left\langle N(y)\right\rangle $ is the average rapidity distribution in a given
centrality class. By definition $\left\langle
a_{i}\right\rangle =0$.

It is straightforward to calculate the two-particle rapidity correlation %
\cite{Bzdak:2012tp} 
\begin{equation}
\frac{C(y_{1},y_{2})}{\left\langle N(y_{1})\right\rangle \left\langle
N(y_{2})\right\rangle }=\left\langle a_{0}^{2}\right\rangle +\left\langle
a_{1}^{2}\right\rangle y_{1}y_{2}+...  \label{eq:C-NN}
\end{equation} 
where\footnote{For clarity we skip $\left\langle a_{0}a_{1}\right\rangle$, 
which vanishes in symmetric (e.g. p+p) collisions.}
\begin{equation}
C(y_{1},y_{2})=\left\langle N(y_{1})N(y_{2})\right\rangle -\left\langle
N(y_{1})\right\rangle \left\langle N(y_{2})\right\rangle .
\end{equation}
As seen in Eq. (\ref{eq:C-NN}), the long-range fluctuation of the fireball rapidity distribution, 
parameterized by fluctuating $a_i$, results in rather nontrivial correlations. The first term
corresponds to a well-known rapidity independent multiplicity fluctuation,
and it can be driven by, e.g., the impact parameter or volume
fluctuation. The second term, $\sim y_{1}y_{2}$, is related to the
fluctuating forward-backward asymmetry in rapidity. In the
wounded nucleon model \cite{Bialas:1976ed,Bialas:2004su} $\langle
a_{1}^{2}\rangle \sim \langle (w_{L}-w_{R})^{2}\rangle ,$
where $w_{L(R)}$ is the number of left(right)-going wounded nucleons \cite{Bzdak:2012tp}.
Recently, the ATLAS Collaboration observed $\langle
a_{1}^{2}\rangle y_{1}y_{2}$ in the two-particle rapidity correlation
functions measured in p+p, p+Pb and Pb+Pb collisions \cite{Aaboud:2016jnr}.
They found that at a given event multiplicity $N_{\mathrm{ch}}$, $%
\langle a_{1}^{2}\rangle $ approximately scales with $1/N_{%
\mathrm{ch}}$ and numerically is very similar for all colliding systems.%
\footnote{%
One would expect rather different results in, e.g., peripheral Pb+Pb
and ultracentral p+p collisions ($N_{\mathrm{ch}} \sim 150$
\cite{Aaboud:2016jnr}). The ATLAS result suggests that the number of
particle sources (at a given $N_{\mathrm{ch}}$) and their fluctuations are
actually similar in all measured systems.} This surprising result still
calls for a quantitative explanation.

\section{Transverse-momentum multiplicity correlations}

It is proposed here to analyze, in a similar way, the rapidity dependence of
transverse momentum and especially transverse-momentum multiplicity correlations. 

Analogously to Eq. (\ref{eq:N(y)}) we have 
\begin{equation}
\frac{P_{t}(y)}{\left\langle P_{t}(y)\right\rangle }=1+b_{0}+b_{1}y+...,
\end{equation}%
where $P_{t}(y)$ is the average (in one event) transverse momentum of
particles in a given rapidity bin $y$
\begin{equation}
P_{t}(y)=\frac{1}{N}\sum\nolimits_{i=1}^{N}p_{t}^{(i)},
\end{equation}
where $N$ is the number of particles (in a given event) at~$y$. Here $%
p_{t}^{(i)}$ is the transverse momentum magnitude of the $i$-th particle. $%
\left\langle P_{t}(y)\right\rangle $ is the average of $P_{t}(y)$ over many
events in a given centrality class. We note that $\left\langle b_{i}\right\rangle =0$, 
in close analogy to the $a_{i}$ coefficients.

The transverse momentum correlation function (studied extensively in the
literature for rather different reasons, see, e.g., \cite%
{Gavin:2006xd,Gavin:2016hmv,Braun:2003fn}) reads 
\begin{equation}
\frac{C_{[P,P]}(y_{1},y_{2})}{\left\langle P_{t}(y_{1})\right\rangle
\left\langle P_{t}(y_{2})\right\rangle }=\left\langle b_{0}^{2}\right\rangle
+\left\langle b_{1}^{2}\right\rangle y_{1}y_{2}+...,  \label{eq:C-PP}
\end{equation}
where 
\begin{equation}
C_{[P,P]}(y_{1},y_{2})\equiv \left\langle
P_{t}(y_{1})P_{t}(y_{2})\right\rangle -\left\langle
P_{t}(y_{1})\right\rangle \left\langle P_{t}(y_{2})\right\rangle.
\end{equation}

The first term in Eq. (\ref{eq:C-PP}) describes an event-by-event rapidity
independent transverse momentum fluctuation. This could be driven for
example by an event-by-event long-range multiplicity fluctuation (if event
multiplicity is correlated with $P_{t}$). The second term describes the
forward-backward rapidity asymmetric transverse momentum fluctuation. 
A possible source of this effect is the forward-backward fireball multiplicity
fluctuation.

It would be especially interesting to measure an event-by-event relation
between $a_{i}$ and $b_{i}$ coefficients. In order to do this, one can
construct a simple correlation function 
\begin{equation}
C_{[N,P]}(y_{1},y_{2})\equiv \left\langle N(y_{1})P_{t}(y_{2})\right\rangle
-\left\langle N(y_{1})\right\rangle \left\langle P_{t}(y_{2})\right\rangle ,
\end{equation}
witch correlates multiplicity and transverse momentum, see, e.g., \cite{Braun:2003fn}. This results in 
\begin{equation}
\frac{C_{[N,P]}(y_{1},y_{2})}{\left\langle N(y_{1})\right\rangle
\left\langle P_{t}(y_{2})\right\rangle }=\left\langle
a_{0}b_{0}\right\rangle +\left\langle a_{1}b_{1}\right\rangle y_{1}y_{2}+...
\end{equation}

The meaning of mixed coefficients $\left\langle a_{i}b_{k}\right\rangle $ is
easy to understand.  
The first term describes the relation between rapidity independent
fluctuation of multiplicity and transverse momentum. The second term is
particularly interesting and describes how rapidity asymmetry in
multiplicity is related to rapidity asymmetry of transverse momentum. If the
particle multiplicity and $P_{t}$ are not correlated then $\left\langle
a_{i}b_{k}\right\rangle =\left\langle a_{i}\right\rangle \left\langle
b_{k}\right\rangle =0$.

In general, the above correlation functions can be expanded in terms of the
orthogonal polynomials \cite{Bzdak:2012tp}. For example 
\begin{equation}
\frac{C_{[N,P]}(y_{1},y_{2})}{\left\langle N(y_{1})\right\rangle
\left\langle P_{t}(y_{2})\right\rangle }=\sum\nolimits_{i,k}\left\langle
a_{i}b_{i}\right\rangle T_{i}(y_{1})T_{k}(y_{2}),
\end{equation}%
with $T_{i}$ being, e.g., the Chebyshev or the Legendre polynomials \cite%
{Bzdak:2012tp,Jia:2015jga}, and analogously for Eqs. (\ref{eq:C-NN}) and (\ref{eq:C-PP}).

\section{Discussion and conclusions}

Several comments are in order.

Consider a set of events with $a_{1}>0$, i.e., the fireball multiplicity is larger
for positive $y$, $N(y) \sim a_{1}y$. The question is what is the rapidity
dependence of the transverse momentum in this case. If $P_{t}$ is also
larger for positive $y$ then $b_{1}>0$ and thus $\langle
a_{1}b_{1}\rangle >0$. This scenario is expected in a typical
hydrodynamical framework, see, e.g., \cite{Bozek:2013sda}.

For example, in the color glass condensate (CGC) framework \cite%
{Gelis:2010nm,Blaizot:2016qgz} one could expect a rather different
conclusion. Consider a proton-proton event, where the two protons are
characterized by different saturation scales, $Q_{1}$ and $Q_{2}$. The
importance of such fluctuations was recently discussed in Refs. \cite%
{Marquet:2006xm,McLerran:2015lta,McLerran:2015qxa,Mantysaari:2016jaz,Bzd-Dus}%
. Here $Q_{1}^{2}=Q_{0,1}^{2}e^{+\lambda y}$ and $Q_{2}^{2}=Q_{0,2}^{2}e^{-%
\lambda y}$ with $\lambda \sim 0.3$, see, e.g., \cite{Praszalowicz:2015dta}.
We choose $Q_{0,1}>Q_{0,2}$ so that in a given rapidity bin, say $|y|<2$, $%
Q_{1}>Q_{2}$, resulting in rapidity asymmetric $N(y)$. 
In this case \cite{Dumitru:2001ux,Bozek:2013sda} 
\begin{eqnarray}
N(y) &\sim &S_{t}Q_{2}^{2}\left[ 2+\ln \left( Q_{1}^{2}/Q_{2}^{2}\right) %
\right] , \\[5pt]
P_{t}(y) &\sim &\frac{2Q_{1}-\frac{2}{3}Q_{2}}{1+\ln \left(
Q_{1}/Q_{2}\right) },
\end{eqnarray}
that is, in CGC the multiplicity is driven by the smaller scale in
contrast to the transverse momentum controlled by the larger one \cite{Dumitru:2001ux}. 
Since $Q_{1}^{2}\sim e^{+\lambda y}$ and 
$Q_{2}^{2}\sim e^{-\lambda y}$, the multiplicity and the transverse momentum
rapidity asymmetries have different signs. If $N(y)$ is growing with
rapidity, then $P_{t}(y)$ is decreasing with $y$. Consequently $a_{1} \gtrless 0$
means $b_{1} \lessgtr 0$ and $\left\langle a_{1}b_{1}\right\rangle <0$. Clearly, this
observation should be treated with caution and more detailed calculations are
warranted, see, e.g., \cite{Duraes:2015qoa,Deja:2017dqh}. The sole purpose of this
exercise was to demonstrate that the sign of $\left\langle
a_{1}b_{1}\right\rangle $ is not at all obvious, and could potentially discriminate 
between different models of particle production. 

As discussed earlier, the ATLAS Collaboration reported a surprising
scaling of $\langle a_{1}^{2}\rangle $ in p+p, p+Pb and Pb+Pb
collisions \cite{Aaboud:2016jnr}. At a given event multiplicity 
$N_{\mathrm{ch}}$, $\langle a_{1}^{2}\rangle $ scales with $1/N_{\rm ch}$ and is
quantitatively very similar for all three systems. It would be
very interesting to see if $\langle b_{1}^{2}\rangle $ and $%
\left\langle a_{1}b_{1}\right\rangle $ satisfy similar scaling.

Obviously, it would be also desired to study higher order
correlation functions \cite{Bzdak:2015dja,DiFrancesco:2016srj}.

An alternative way to analyze the above correlation functions is 
the principal component analysis, discussed in Ref. \cite{Bhalerao:2014mua}.

In conclusion, it is proposed to analyze the rapidity dependence of
transverse momentum and in particular transverse-momentum multiplicity
correlation functions using the orthogonal polynomial
expansion. A careful study of the coefficients $\langle
a_{i}^{2}\rangle ,$ $\langle b_{i}^{2}\rangle $ and $\langle a_{i}b_{k}\rangle $ 
could potentially discriminate between different 
models of particle production, and reveal detailed information on the longitudinal 
structure of systems created in p+p, p+A and A+A collisions.

\bigskip

\vspace{\baselineskip} 
\noindent\textbf{Acknowledgments} 
\newline
{} 
We thank Piotr Bo\.zek and Volker Koch for useful comments. 
This work is supported by the
Ministry of Science and Higher Education (MNiSW) and by the National Science
Centre, Grant No. DEC-2014/15/B/ST2/00175, and in part by
DEC-2013/09/B/ST2/00497.

\end{document}